\documentclass{rspublic}

\usepackage{amsmath}
\usepackage{amsfonts}
\usepackage{amsthm}
\usepackage{amscd}

\newcommand{\Rr}{{\mathbb R}}
\newcommand{\Cc}{{\mathbb C}}
\newcommand{\Zz}{{\mathbb Z}}
\newcommand{\half}{{\textstyle \frac{1}{2}}}

\newcommand{\mvec}{{\bf m}}
\newcommand{\nvec}{{\bf n}}
\newcommand{\pvec}{{\bf p}}

\newcommand{\Avec}{{\bf A}}
\newcommand{\rvec}{{\bf r}}

\newcommand{\thetavec}{{\boldsymbol \theta}}

\newcommand{\Phivec}{{\boldsymbol \Phi}}
\newcommand{\omegavec}{{\boldsymbol \omega}}

\newcommand{\Ccal}{{\cal C}}

\newcommand{\Hcal}{{\cal H}}

\newcommand{\grad}{{\boldsymbol\nabla}}

\newcommand\ket[1]{|#1\rangle}
\newcommand\bra[1]{\langle #1|}
\newcommand\braket[2]{\langle #1| #2 \rangle}

\newcommand{{\ad}}{\,\text{ad}}
\usepackage{color}
\usepackage{graphics}
\usepackage{subfigure}

\newcommand{\Gbar}{{\overline G}}
\newcommand{\Hbar}{{\overline H}}
\newcommand{\Cbar}{{\overline C}}
\newcommand{\Abar}{{\overline A}}
\newcommand{\Hcalbar}{{\overline {\cal H}}}
\newcommand{\Omegahat}{{\hat \Omega}}

\usepackage{graphicx}

\begin{document}

\title[Quantum statistics on graphs]{Quantum statistics on graphs}
\author[J.M.~Harrison, J.P.~Keating \& J.M.~Robbins]{J.M.~Harrison$^1$, J.P.~Keating$^2$ \& J.M.~Robbins$^2$}
\affiliation{$^1$ Department of Mathematics, Baylor University,
One Bear Place,
Waco, TX 76798-7328
USA\\
$^2$ School of Mathematics,
 University of Bristol, University Walk, Bristol BS8 1TW, UK
} \label{firstpage} \maketitle

%
%
%

\begin{abstract}{Quantum statistics, quantum graphs, anyons}
Quantum graphs are commonly used as models of complex quantum
systems,  for example molecules, networks of wires, and states of
condensed matter. We consider quantum statistics for
indistinguishable spinless particles on a graph, concentrating on
the simplest case of abelian statistics for two particles.  In
spite of the fact that graphs are locally one-dimensional, anyon
statistics emerge in a generalized form.  A given graph may
support a family of independent anyon phases associated with
topologically inequivalent exchange processes. In addition, for
sufficiently complex graphs, there appear new discrete-valued
phases.  Our analysis is simplified by considering combinatorial
rather than metric graphs -- equivalently, a many-particle
tight-binding model. The results demonstrate that graphs provide
an arena in which to study new manifestations of quantum
statistics. Possible applications include topological quantum
computing,  topological insulators, the fractional quantum Hall
effect, superconductivity and molecular physics.
\end{abstract}

\section{Introduction}\label{sec: intro}

The quantum mechanical properties of a many-particle system depend
profoundly on whether the particles are distinguishable or not,
even if, in the former case, the distinguishing properties have
little effect on the underlying classical mechanics. For this
reason, the description of indistinguishable particles in quantum
mechanics -- quantum statistics -- continues to be revisited, both
to gain insight into its foundations (Duck \&  Sudarshan 1997,
Berry 2008) as well as to predict and understand new phenomena,
including the fractional quantum Hall effect (Wilczek 1990, Jain
2007), superconductivity (Wilczek 1990), topological quantum
computing (Nayak {\it et al}.~2008) and particle-like states of
nonlinear field theories, where standard quantization procedures
are not straightforward to apply (Finkelstein \& Rubenstein 1968,
Manton 2008).

The standard treatment of quantum statistics takes as its starting point the
quantum mechanical description of a single particle.  The $n$-particle Hilbert
space, $\Hcal_n$, is then taken to be the tensor product of $n$ copies of the
one-particle Hilbert space, $\Hcal$.  The fact that the particles are
indistinguishable means that physically observable operators commute with
permutations of the particle labels.   This implies that $\Hcal_n$ may
be
decomposed into physically distinct components characterized by different permutation symmetries.
The symmetrization postulate restricts the physically realizable spaces to
either the symmetric component, which obeys Bose statistics, or the
antisymmetric component, which obeys Fermi statistics; components characterized
by more complicated behaviour under permutations -- parastatistics -- are thus
excluded.  Finally, the spin-statistics relation determines which of the two
allowed alternatives applies -- Bose statistics for integer spin, and Fermi
statistics for half-odd-integer spin.

There is another approach to quantum statistics which is based on the topology
of the classical configuration space. It takes as its starting point the
configuration space, $M$, of a single particle. The $n$-particle configuration
space, $C_n(M)$, is then taken to be the cartesian product of $n$ copies of the
one-particle configuration space, with, crucially, coincident configurations
excluded (no two particles can be in the same place) and permuted
configurations identified.  So defined, $C_n(M)$ may have nontrivial topology.
In particular, curves along which the particles are exchanged (some particles
returning to where they started, others ending up where another started) are
regarded as closed curves (as permuted configurations are identified), and
these exchange curves cannot be continuously contracted to a single point.  If
such curves exist, then $C_n(M)$ is multiply connected.

Quantum mechanics on a multiply connected configuration space $\Ccal$, whether
$\Ccal$ describes $n$ particles or just one, allows for an additional freedom
in the quantum description, namely the choice of a representation of the
fundamental group, $\pi_1(\Ccal)$, of $\Ccal$ . The (single-particle)
Aharonov-Bohm effect provides a familiar example: the configuration space
$\Ccal$ for a particle with charge $q$ in the presence of an impenetrable
solenoid, say along the $z$-axis, may be taken to be $\Rr^3$ with the solenoid
-- an infinite cylinder -- removed.  $\Ccal$ is multiply connected; closed
curves may be classified according to the number of times they wind around the
excluded cylinder. To the canonical momentum operator, $\pvec = -i\nabla$, we
can add a vector potential, $\Avec$, whose curl vanishes away from the
cylinder, so that the commutation relations are preserved, as are the
classical equations of motion -- there are no physically accessible magnetic
fields. However, the line integral of $\Avec$ around the cylinder need not
vanish, and may be realized by a magnetic flux $\phi$ inside the cylinder,
measured in units of the flux quantum $qh/c$.  The flux determines a
representation of $\pi_1(\Ccal)$ that assigns to a closed curve with $m$
windings the phase factor $\exp(im\phi)$.  The value of the flux modulo $2\pi$
-- equivalently, the choice of a representation of $\pi_1(\Ccal)$ -- affects
the quantum mechanical properties, although observable consequences vanish in
the classical limit.

More generally, on a multiply-connected configuration space $\Ccal$, the
momentum is described by a covariant derivative which, in the absence of
external fields, has zero curvature (so that the classical mechanics is
unaffected) but which may engender nontrivial holonomy.  The holonomy corresponds to a
representation of the fundamental group $\pi_1(\Ccal)$. The representation
could be one-dimensional and therefore abelian, assigning (commuting) phase
factors to noncontractible closed curves, or higher-dimensional, assigning
$d\times d$ unitary matrices (in general, non-commuting)  to noncontractible closed curves.
The unitary matrices act on the wavefunctions, which are
taken to have $d$ components (more generally, wavefunctions are taken to be
sections of a $d$-dimensional complex vector bundle).

Taking $\Ccal$ to be the configuration space $C_n(M)$ of $n$
indistinguishable particles, Leinaas and Myrheim (1977) showed
that representations of the fundamental group $\pi_1(C_n(M))$
determine the possible realizations of quantum statistics. Similar
conclusions were reached by Laidlaw and de Witt (1971) using path
integrals, an approach that has been applied in a variety of
settings, including cases where canonical quantization is
difficult to carry out (Finkelstein and Rubenstein 1968,
Balachandran {\it et al}.~1993, Manton 2008).

If $\pi_1(C_n(M))$ coincides with the symmetric group, $S_n$, then
the topological approach to quantum statistics yields the same
possibilities as does the standard one, namely Bose or Fermi
statistics, or, if one gives up the symmetrization postulate,
parastatistics, which are associated with higher-dimensional
representations of $S_n$. This is the case for $M = \Rr^d$ with $d
> 2$. As is well known, the situation is  different for $M =
\Rr^2$, i.e.~for particles in the plane.  The fundamental group
$\pi_1(C_n(\Rr^2))$ is the braid group $B_n(\Rr^2)$, which has
infinitely many elements.  The characterization of unitary
representations of the braid group is not complete (see, e.g.,
Birman \& Brendle 2004), but its one-dimensional (abelian)
representations include a new type of statistics -- anyon
statistics (Wilczek 1990). Closed curves on which two of the
particles wind around each other $m/2$ times (odd values of $m$
correspond to an exchange of positions) are assigned a phase
factor $\exp (im \alpha)$, independently of any phases associated
with external gauge fields. The special values $\alpha = 0$ and
$\alpha = \pi$ correspond to Bose and Fermi statistics
respectively.

Anyon statistics have been found to provide deep insight into phenomena
involving strongly interacting electrons, for
 example the fractional quantum
Hall effect (Laughlin 1983). In the composite fermion theory (Jain
2007), for example, the collective excitations of the
many-electron system are approximately described by an effective
Hamiltonian in which they obey anyon statistics with phase
$\alpha$ related to the fractional Hall conductance.

The situation is different again for $M = \Rr$, i.e.~particles on the line.  Here
the fundamental group $\pi_1(C_n(\Rr^2))$ is trivial; there are no
noncontractible closed curves, as it is not possible to exchange particles
without one passing through another.  Thus, the topological approach predicts
only Bose statistics (but see Leinaas and Myrheim (1977) for yet another approach to
quantum statistics in one dimension).

One motivation for this paper is to study quantum statistics for
particles on networks of one-dimensional wires, or metric graphs.
A metric graph $\Gamma$ is a collection of nodes, or vertices,
connected by edges, or one-dimensional intervals, of specified
lengths. Single-particle quantum mechanics on  metric graphs has
been extensively studied. The model was originally introduced to
describe free electrons on chemical bonds (Pauling 1936). Some
current applications include superconductivity (de Gennes 1981),
quantum chaos (Kottos and Smilansky 1997, Keating 2008), and
Anderson localization (Aizenman {\it et al.} 2006).
The particle is described by a set of wavefunctions, $\Psi_\epsilon(x_\epsilon)$, one
for each edge of $\Gamma$.  Each wavefunction 
is acted on by a Hamiltonian, $H_\epsilon$. For example,
$H_\epsilon$ could be a Schr\"odinger operator with a gauge
potential, i.e.~of the form 
$\half
(-id/dx_\epsilon - A_\epsilon(x_\epsilon))^2 + V_\epsilon(x_\epsilon)$.

The boundary conditions at the vertices of the quantized graph is
a subtle point in the theory. For free particles, i.e.~$H_\epsilon
= -\half d^2/dx_\epsilon^2$, a simple choice motivated by physical
considerations is to require that i) wavefunctions are continuous
across vertices and ii) the sum of the outgoing derivatives at
each vertex should vanish (Neumann-like boundary conditions). A
complete characterization of boundary conditions that render the
free-particle Hamiltonian self-adjoint has been given in terms of
the Lagrangian planes of a certain complex symplectic boundary
form (Kostrykin and Schrader 1999, Kuchment 2004). For more
general operators, the classification of self-adjoint boundary
conditions remains an open problem (see Bolte and Harrison (2003)
for a corresponding classification scheme for the Dirac operator).
For a given graph model, one can in principle derive physically
appropriate boundary conditions from the limiting behaviour of a
more realistic two- or three-dimensional model, but the analysis
can be quite involved.

As a metric graph is locally one-dimensional, one might imagine
that quantum statistics on a metric graph would be limited to Bose
or Fermi statistics. This turns out not to be the case, as was
previously demonstrated in particular examples by Balachandran \&
Ercolessi (1992) (discussed in
Section~\ref{sec: examples}\ref{subsec: lasso and
figure-of-eight}).
However, a systematic treatment of quantum statistics for general metric
graphs has not been given.  One difficulty is that the quantization of
many-particle metric graphs poses analytical challenges beyond those
encountered in the one-particle theory.  Besides boundary conditions for a
single particle at a vertex, one has also to specify boundary conditions for
coincidences of two or more particles on edges and at vertices that render
the many-particle Hamiltonian self-adjoint.

Here we circumvent these 
difficulties while preserving the essential topological features
by considering the simpler problem of indistinguishable particles
on a {\it combinatorial graph} $G$.  A combinatorial graph
consists of a set of vertices together with a specification of
pairs of vertices that are connected by edges.  The edges
themselves do not contain any points, so the configuration space
for the particle consists of the vertices alone, and hence is
zero- rather than one-dimensional. Quantum mechanics on a
combinatorial graph is a tight-binding model on the vertices.
Thus, quantum statistics on combinatorial graphs is an interesting
problem in its own right.

It turns out that the $n$-particle configuration space, $C_n(G)$,
can also be regarded as a combinatorial graph, and hence, in
contrast to a many-particle metric graph, is easily quantized.
Following the topological approach, we can characterize the
possible quantum statistics in terms of representations of the
fundamental group of $C_n(G)$.  As $C_n(G)$ is a discrete space,
it is not immediately apparent what is meant by its fundamental
group, but  one can define a combinatorial version that
coincides with its
metric-graph counterpart. 

For simplicity, we consider here two particles on a combinatorial
graph, and restrict our attention to abelian (one-dimensional)
representations of the fundamental group.    
It turns out that a given graph may support a family of
independent anyon phases associated with topologically
inequivalent exchange processes. In addition, for sufficiently
complex graphs, there appear new discrete-valued phases.
(Mathematically, these discrete phases correspond to
representations of the torsion component of the abelianized
fundamental group, or, equivalently, the first homology group of
$C_n(G)$.) The results extend straightforwardly to abelian
statistics for $n
> 2$ particles. The extension to nonabelian statistics will be
discussed in a forthcoming paper.

The paper is organized as follows. In Section~\ref{sec:
one-particle graph} we present quantum mechanics on a
combinatorial graph as a tight-binding model. We introduce gauge
potentials, the analogue of vector potentials, which assign phase
factors to the edges of the graph.  These phase factors are then
incorporated into the off-diagonal matrix elements (transition
amplitudes) of
the Hamiltonian matrix. 
Gauge potentials are determined, up to a
choice of gauge, by the accumulated phases along, or fluxes through, the cycles of the
graph.

In Section~\ref{sec: 2 particles} we consider two
indistinguishable particles on a combinatorial graph. The
two-particle configuration space can itself be regarded as a
combinatorial graph (usually larger than the original graph), that
can be quantized following the general prescription of
Section~\ref{sec: one-particle graph}.  We characterize, up to a
choice of gauge, {\it topological gauge potentials} on the
two-particle graph, which determine the quantum statistics.
Topological gauge potentials have the property that the only
cycles with nonzero flux  correspond to nontrivial closed curves
in the metric setting. They are parameterized by a set of {\it
free statistics phases}, which range between $0$ and $2\pi$, as
well as, in some cases, a set of {\it discrete statistics phases}
whose values are constrained to certain rational multiples of
$2\pi$. A certain subset of the free statistics phases may be
attributed to Aharonov-Bohm flux lines threading the one-particle
graph; the remainder describe many-body effects
(Section~\ref{sec: 2 particles}\ref{subsec: AB potentials}).
Bose and Fermi
statistics, as understood in the conventional approach, may be recovered from
particular choices of topological gauge potential
(Section~\ref{sec: 2 particles}\ref{subsec: Bose and Fermi})
Our results are
illustrated by a number of examples in Section~\ref{sec: examples}, and in the
concluding remarks we consider perspectives for further investigations and
possible applications.

\section{Quantum mechanics on combinatorial graphs}
\label{sec: one-particle graph}
\subsection{Combinatorial graphs}\label{subsec: 1-particle comb graph}
A combinatorial graph $G$ consists of a set $V=\{1,\ldots,v\}$ of sites, or
vertices, labeled $1$ through $v$, which may be connected by bonds, or edges.
It is convenient to describe the  connectivity of the graph, i.e.~its edges, by a
$v \times v$ {\it adjacency matrix}, $A$, whose $(j,k)^{th}$ entry gives the
number of edges from vertex $j$ to vertex $k$. We assume that $G$ is {\it
undirected}, i.e.~$A_{jk} = A_{kj}$.
We write $j \sim k$ to indicate that $j$ and $k$ are connected by an edge. The
number of edges in the graph, $e$, is then given by $e = \half \sum_{jk} A_{jk}$.
Edges will be labeled by Greek indices which take values between $1$
and $e$. Given an edge $\epsilon$, we let $\epsilon_<$ and
$\epsilon_>$ denote its vertices of lower and higher index
respectively. Sometimes it will be useful to assign an orientation
to an edge; our convention will be that  positive orientation
corresponds to going from $\epsilon_<$ to $\epsilon_>$.

We also assume that $G$ is {\it simple}, so that there is at most one edge
between any two  vertices, and that no vertex is connected to itself,
i.e.~$A_{jk}$ is equal to zero or one, and $A_{jj}$ is equal to zero.
Any graph can be made simple by introducing additional vertices on its
multiply-connecting and  self-connecting edges. The valency, or {\it degree},
$v_j$, of the vertex $j$ is the number of edges connected to $j$, i.e.~$v_j = \sum_{k} A_{jk}$.
Finally, we assume that $G$ is {\it connected}, i.e.~every pair of
vertices is connected by some sequence  of edges.  In terms of the
adjacency matrix, this means that, for any given vertices $j$ and
$k$, $(A^n)_{jk} \neq 0$ for some $n$.

While our focus is on combinatorial graphs, some aspects of
our treatment may be motivated by considering {\it metric
graphs} associated to a combinatorial graph. Given
 a combinatorial graph, $G$, we can associate to it a  metric
graph, $\Gamma$, by assigning a length $L_\epsilon>0$ to each edge
$\epsilon$ of $G$.  On $\Gamma$, $\epsilon$ is
regarded as an interval $[0,L_\epsilon]$, with $0$ identified with
$\epsilon_<$ and $L_\epsilon$ identified with $\epsilon_>$.   One
can define continuous curves on $\Gamma$, and a metric is obtained
by taking the distance between two points to be the minimum length
of continuous curves joining the points. One can consider closed
curves based at a vertex $*$ and define the fundamental group
$\pi_1(\Gamma,*)$ in the usual way.

Next we recall some basic facts about the topology of combinatorial graphs (see, e.g.,
Hatcher 2001),
which will play a role in their quantization.
A {\it path} $p = (j_0, j_1, \ldots, j_n)$ on a combinatorial graph
$G$ is a sequence of vertices in which consecutive vertices are
connected by edges, i.e.~$j_r \sim j_{r+1}$ (thus, consecutive
vertices must be distinct). The length of a path is the number of
edges along it. 
A single vertex, $j$, may be regarded as a path of zero length, and an
edge, $\epsilon$, regarded as the path $(\epsilon_<,\epsilon_>)$.
Given a path $p = (j_0, j_1, \ldots, j_n)$, we define the inverse of
$p$, denoted $p^{-1}$, to be the path $(j_n, j_{n-1}, \ldots, j_0)$.
If $p = (j_0,\ldots,j_n)$ and $q = (k_0,\ldots,k_r)$ are two paths
such that the last vertex of $p$ coincides with the first vertex of
$q$, 
we define the concatenation, or product,
of $p$ and $q$, denoted $pq$, to be the path $(j_0, \ldots, j_n,
k_1,\ldots, k_r)$.

The path $qpp^{-1}r$ describes the product of paths $q$ and $r$
with an intervening retracing of the path $p$. We want to regard
paths that differ by retracings of intermediate components as
being the same. Thus, we introduce the equivalence relation
$   q p p^{-1} r \equiv qr$.
Paths that are equivalent to a zero-length path, i.e.~to their initial vertex,
are called {\it self-retracing}.

A {\it cycle} is a path $c = (j_0, j_1, \ldots, j_n)$ that is
closed, i.e.~$j_n = j_0$.  A cycle is {\it primitive} if its vertices,
apart from the first and last, are all distinct. The set of cycles
on a combinatorial graph can be characterized with the aid of a {\it
spanning tree}, which we define next. A {\it tree} is a combinatorial graph whose only cycles
are self-retracing.  A {\it subgraph} of a combinatorial graph $G$
is a combinatorial graph whose vertices and edges are subsets of the
vertices and edges of $G$.  A spanning tree, $T$, of
$G$ is a subgraph that is a connected tree containing all
the vertices of $G$.
%
Thus, given any two vertices of $G$, there is a path on $T$,
unique up to retracings, that joins them. $G$ has at least one
spanning tree, and, unless $G$ is itself a tree, more than one.
(An iterative algorithm for constructing a spanning tree is to
remove an edge from a primitive cycle of $G$ until no primitive
cycles remain).  Clearly, any spanning tree of $G$ has $v-1$
edges. We let $f$ denote the number of edges of $G$ not in a
spanning tree, so $f$ is one minus the Euler characteristic of
$G$,
\begin{equation}\label{eq: f for G}
    f = e - (v-1).
\end{equation}

Let $T$ denote a spanning tree of $G$ and $*$ a vertex of $G$.  Let
us label the edges of $G$ that are not in $T$ by an index $\phi$,
$1 \le \phi \le f$. Let $c_\phi(*)$ denote the cycle obtained by
proceeding along the (unique) path with no retracings from $*$ to
$\phi_<$ on $T$, then from $\phi_<$ to $\phi_>$ along $\phi$, and
finally from $\phi_>$ back to $*$ along the (unique) path with no
retracings on $T$. We call $c_\phi(*)$ a {\it fundamental cycle}.
The definition depends on the choice of $T$ and $*$, but the
dependence on $*$ is easily accounted for, as
\begin{equation}\label{eq: c(**_}
    c_\phi(**) \equiv p c_\phi(*) p^{-1},
\end{equation}
where $p$ is the (unique) path on $T$ from $**$ to $*$. An
arbitrary cycle $c$ beginning and ending at $*$ can be expressed,
up to retracings, as a product of fundamental cycles and their
inverses, i.e.
\begin{equation}\label{eq: arbitrary cycle}
    c \equiv c^{s_1}_{\phi_1}(*) \cdots c^{s_t}_{\phi_t}(*),
\end{equation}
where $s_j = \pm 1$.
(More formally,
the set of cycles based at $*$ modulo
retracings forms a group, the combinatorial fundamental group $\pi_1^C(G,*)$.
$\pi_1^C(G,*)$ is a free group on $f$ generators $c_1(*), \ldots, c_f(*)$ and is
isomorphic to the fundamental group,  $\pi_1(\Gamma,*)$, of a metric graph, $\Gamma$,
associated to $G$.)

\subsection{Quantization}\label{subsec: 1-particle quantization}

We regard the set of vertices $V$ of a combinatorial graph $G$ as a
configuration space. $V$ might be the configuration space of a
single particle, but we shall not restrict ourselves to this point of view.
Indeed, in Section~\ref{sec: 2 particles} we will consider combinatorial graphs whose
vertices represent configurations of two particles.
We take quantum mechanics on $G$ to be given by a tight-binding
model on the set of vertices, in which short-time transitions are
allowed only between vertices connected by edges. The Hilbert
space is $\Cc^v$, with basis vectors $\ket{j}$, $1 \le j \le v$,
describing states in which the system is localized at one of the
vertices. Dynamics is given by the Schr\"odinger equation, $
i\dot{\ket{\psi}} = H\ket{\psi}$,
where the  Hamiltonian $H$ is a $v \times v$ hermitian matrix with nonzero
off-diagonal entries only between connected vertices, i.e.
\begin{equation}\label{eq: constraint on H}
    H_{jk} = 0 \text{ if } j \neq k \text{ and } j\nsim k.
\end{equation}
One example of a Hamiltonian is (minus) the {\it combinatorial
Laplacian}, or the discrete kinetic energy, $H = A - D$,
where $D$ is given by $D_{jk} = v_j \delta_{jk}$.

\subsection{Gauge potentials}\label{subsec: gauge potentials}

A {\it gauge potential} $\Omega$ on $G$ is a $v \times v$ real
antisymmetric matrix such that $\Omega{jk}$ vanishes if $j \nsim
k$.
Given a path $p = (j_0, \ldots, j_n)$ on $G$, we define
\begin{equation}\label{eq: flux along p}
    \Omega(p) =  \sum_{r = 0}^{n-1} \Omega_{j_r j_{r+1}}.
\end{equation}
Clearly, if $p$ and $p'$ differ by retracings, then
\begin{equation}\label{eq: Omega and retracings}
    \Omega(p) = \Omega(p'),
\end{equation}
and if $p$ and $q$ can be concatenated, then
\begin{equation}\label{eq: Omega through product of paths}
    \Omega(p q) = \Omega(p) + \Omega(q). 
\end{equation}
For a cycle $c$, we refer to $\Omega(c)$ as the {\it flux} of $\Omega$ through
$c$.  From \eqref{eq: c(**_} and \eqref{eq: Omega and retracings}, it follows
that the flux through a fundamental cycle $c_\phi(*)$ is independent of $*$;
hence we write $\Omega(c_\phi)$, omitting $*$ from the notation.
For an arbitrary cycle $c$, it follows from \eqref{eq: Omega through product of
paths} that
\begin{equation}\label{eq: Omega for general c}
    \Omega(c) = \sum_{j = 1}^r (-1)^{s_j} \Omega(c_{\phi_j}),
\end{equation}
where $\phi_j$ and $s_j$ are given by \eqref{eq: arbitrary cycle}.
Thus,  all fluxes are  determined by the fluxes through a set of fundamental
cycles.

We incorporate a gauge potential $\Omega$ into a Hamiltonian $H$ by multiplying
the transition amplitudes from $j$ to $k$ by the phase factors $\exp\left(i
\Omega_{jk}\right)$. The new Hamiltonian is given by
\begin{equation}\label{eq: incorporate gauge potential}
    H^\Omega_{jk} = H_{jk} \exp\left(i \Omega_{jk}\right)
\end{equation}
Clearly, $H^\Omega$ is hermitian and satisfies (\ref{eq: constraint on H})
(see Oren {\it et al.}~2009 for a similar construction).

We can motivate the prescription \eqref{eq: incorporate gauge potential} by
making the following analogy with vector potentials on a metric graph.
We note that the non-diagonal part of the Hamiltonian can be expressed as a
real linear combination of the $v(v-1)$ transition Hamiltonians
\begin{equation}\label{eq: S,A defined}
    S_{(j,k)} = \ket{j} \bra{k} + \ket{k} \bra{j}, \quad
    A_{(j,k)} = i\left(\ket{j} \bra{k} - \ket{k} \bra{j}\right),
\end{equation}
where $j \sim k$, and, for definiteness, we take $j < k$.
Let $\epsilon$ denote the edge between $j$ and $k$, and consider a metric graph
where $\epsilon$ corresponds to the interval $[0,L_\epsilon]$.  Let
$\Psi_\epsilon(x_\epsilon)$ denote the wavefunction along $\epsilon$ on the quantized metric
graph, and let us formally regard
$\ket{j}$ and $\ket{k}$ as vertex states localized at $x_\epsilon = 0$ and
$x_\epsilon = L_\epsilon$ respectively, so that $\braket{j}{\Psi_\epsilon} =
\Psi_\epsilon(0)$ and $\braket{k}{\Psi_\epsilon} =
\Psi_\epsilon(L_\epsilon)$.  Then, formally, $\ket{k} = T_\epsilon
\ket{j}$, where $T_\epsilon
= \exp(-i p_\epsilon 
L_\epsilon)$ is the unitary translation by a distance 
$L_\epsilon$ along $\epsilon$ generated by the momentum  operator
$p_\epsilon = -i d/dx_\epsilon$. Thus we can rewrite \eqref{eq:
S,A defined} as
\begin{equation}\label{eq: S_jk, A_jk in terms of projector}
    S_{(j,k)} = P_j T^\dag_\epsilon
    + T_\epsilon
    P_j ,
    \quad
    A_{(j,k)} = i \left(P_j T^\dag_\epsilon
    - T_\epsilon
    P_j\right),
\end{equation}
where $P_j$ denotes the projector $\ket{j}\bra{j}$.
Now suppose we introduce a vector potential on the edge $\epsilon$, replacing
$p_\epsilon$ by $p_\epsilon - A_\epsilon(x_\epsilon)$, and make the
substitution
\begin{equation}\label{eq: translation with vector potential}
    T_\epsilon
    \rightarrow \exp \left(-i \int_0^{L_\epsilon} (p_\epsilon -
    A_\epsilon(x_\epsilon))\, dx_\epsilon\right) = e^{i\Omega_{jk}} T_\epsilon,
\end{equation}
where $\Omega_{jk}$ is the integral of $A_\epsilon$ along $\epsilon$.
Making this same substitution in \eqref{eq: S_jk, A_jk in terms of projector}, we obtain
expressions for the transformed transition Hamiltonians,
$S_{(j,k)}^{\Omega}$ and $A_{(j,k)}^{\Omega}$, that are  equivalent to the
prescription \eqref{eq: incorporate gauge potential}.

A  gauge potential is {\it trivial} if it is of the form
\begin{equation}\label{eq: trivial gauge potential}
    \Omega_{jk} = \begin{cases}
    \theta_k - \theta_j + 2\pi M_{jk}, & j \sim k,\\
    0,& \text{otherwise},
    \end{cases}
\end{equation}
where $\thetavec$ is a $v$-tuple of phases and $M$  an (antisymmetric) integer
matrix.
For a trivial gauge potential,
$H^\Omega$ is given by $H^\Omega = U H U^\dag$,
where $U$ is the diagonal unitary matrix $U_{jk} = \exp(i \theta_j)
\delta_{jk}$
that generates the gauge transformation  $\ket{j} \mapsto \exp(i \theta_j)
\ket{j}$. 
The terminology is in analogy with quantum
mechanics on $\Rr^3$, where a vector potential $\Avec(\rvec)$ is trivial if it is a
gradient $\grad \theta$, in which case it is induced by the gauge
transformation $\Psi(\rvec) \mapsto e^{i\theta(\rvec)} \Psi(\rvec)$.

In analogy with the fact that a trivial vector potential has vanishing curl,
we
have the following characterization of trivial gauge potentials on $G$
(we omit the argument, which is not difficult):
\begin{equation}\label{eq: used to be prop}
\text{$\Omega$ is trivial if and only if $\Omega(c)$ is an integer
multiple of 2$\pi$ for every cycle $c$.}
\end{equation}
It follows from \eqref{eq: used to be prop}
and \eqref{eq: Omega for general c}
that a gauge potential $\Omega$ is
determined up to a gauge transformation by its fluxes through a set of
fundamental cycles. Indeed, given a set of fluxes $\omega_\phi$, we
can construct a gauge potential $\Omega$ for which $\Omega(c_{\phi}) =
\omega_\phi$, as follows: if $j$ and $k$ are vertices of an edge in $T$, we
take
\begin{equation}\label{eq: Omega_jk on tree}
    \Omega_{jk} = \Omega_{kj} = 0.
\end{equation}
For an edge $\phi$ not in $T$ with vertices $j = \phi_<$ and $k = \phi_>$, we
take
\begin{equation}\label{eq: Omega given omega}
\Omega_{jk} = -\Omega_{kj} = \omega_{\phi}.
\end{equation}
%

\section{Two particles }\label{sec: 2 particles}
\subsection{Quantization}\label{subsec: 2 particle quantization}
Given a one-particle configuration space, $X$, there is a standard construction
for the configuration space,  $C_2(X)$, of two indistinguishable particles on $X$ (the
construction generalizes straightforwardly to more than two particles).
$C_2(X)$ consists of unordered pairs of distinct
points of $X$, i.e.
\begin{equation}\label{eq: two-particle configuration space}
    C_2(X) = \{X \times X - \Delta_2(X)\}/S_2,
\end{equation}
where $\Delta_2(X) = \{(x,x)\}$ 
denotes the coincident
two-particle configurations, which are excluded, and $S_2$ denotes the
symmetric group, whose single nontrivial element -- exchange -- acts 
on
$X\times X$ 
according to $(x,y) \mapsto (y,x)$.

Given a combinatorial graph $G$, we shall now regard the set of its vertices, $V
= \{1,\ldots, v\}$, as a one-particle configuration space. Hence, we take
$G_2 := C_2(V)$ to be the configuration space for two indistinguishable particles on $G$.
Depending on context, we denote configurations in $G_2$ either by unordered
pairs of vertices $\{j,k\}$, or by ordered pairs $(j,k)$ with $j < k$.

We may regard  $G_2$ 
as a new combinatorial graph.
Nodes (or $G_2$-vertices) $(j,l)$ and $(k,m)$ are taken to be connected by an
edge if they have one $G$-vertex in common
while their other $G$-vertices are connected by an edge of $G$.
Equivalently, moving along an edge of $G_2$ corresponds to keeping one particle
fixed at a vertex of $G$ while moving the other along an edge of $G$. The
adjacency matrix of $G_2$, denoted $A_{2}$, is given by
\begin{equation}\label{eq: A^(2)}
    A_{2; (j,l),(k,m)} = \delta_{jk} A_{lm} + \delta_{jm} A_{lk} +
    A_{jk} \delta_{lm} +  A_{jm} \delta_{lk},
\end{equation}
where $A$ is the adjacency matrix of $G$.  (To avoid awkward notation, we will
not relabel the vertices of $G_2$ by a single index.)
The number of  vertices and edges of $G_2$, denoted $e_2$ and $v_2$
respectively, are given by
\begin{equation}\label{eq: v_2 and e_2}
    v_2 = v(v-1)/2, \quad e_2 = e(v-2).
\end{equation}

We  define quantum mechanics on $G_2$  following the prescription of
Section~\ref{sec: one-particle graph}\ref{subsec: 1-particle
quantization}.
The Hilbert space is
$\Cc^{v_2}$ with basis vectors $\ket{jl}$, $j < l$, representing states
where one particle is at site $j$ and the other at site $l$. Two-particle
Hamiltonians, denoted $H_2$, are $v_2$-dimensional hermitian matrices
$H_{2;jl,km}$, where $j < l$ and $k < m$, whose off-diagonal elements, i.e., elements for
which $(j,l) \neq (k,m)$, vanish whenever $A_{2; (j,l),(k,m)}$ vanishes.
Thus,  short-time transitions generated by $H_2$ involve one particle making
an allowed transition on $G$ while the other particle remains fixed.

Given a
one-particle Hamiltonian $H$ on $G$, we can construct a two-particle
Hamiltonian $H_2^\sigma$ on $G_2$ according to
\begin{equation}\label{eq: H^2}
    \bra{jl} H_2^\sigma\ket{km} = \delta_{jk} H_{lm}+
    \sigma \delta_{jm} H_{lk} + \sigma \delta_{lk} H_{jm} +
    \delta_{lm} H_{jk}, \quad j < l, k < m,
\end{equation}
where $\sigma = \pm 1$ (one can check that $H_2^\sigma$ satisfies
(\ref{eq: constraint on H})).  As discussed in
Section~\ref{sec: 2 particles}\ref{subsec: Bose and Fermi}
below, $\sigma = -1$  describes noninteracting fermions, while
$\sigma = +1$ describes hard-core (and therefore interacting)
bosons which are prevented from occupying the same site.  A simple
illustration (two free particles on a linear graph) is given in
Section~\ref{sec: examples}\ref{subsec: linear}.

%

\subsection{Topological gauge potentials}\label{subsec: 2 particle topological gauge potentials}

Gauge potentials can be introduced on a two-particle combinatorial graph following the
general prescription of
Section~\ref{sec: one-particle graph}\ref{subsec: gauge
potentials}.
Quantum statistics is described
by a subset of these, which we call {\it topological gauge potentials}.
As discussed below, topological gauge potentials
correspond to gauge potentials on a two-particle metric graph whose
fluxes through all contractible closed curves vanish (modulo $2\pi$).  Thus, their effects
are purely quantum mechanical, vanishing in the classical limit.

\begin{figure}[htb]
\begin{center}
  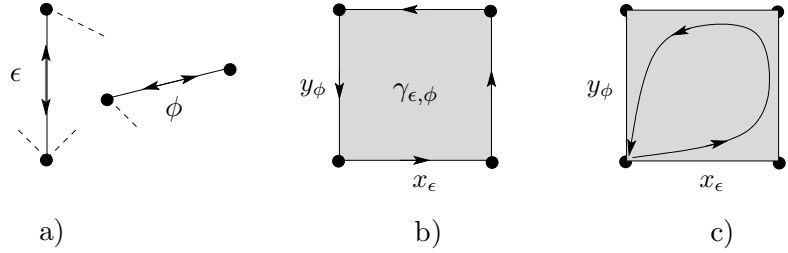
  \caption{a)  The cycle $c_{\epsilon,\phi}$  and b)  corresponding loop $\gamma_{\epsilon,\phi}$
  in the two-particle configuration space of the metric graph $\Gamma$.  c) Contraction of $\gamma_{\epsilon,\phi}$}
  \label{fig: contractible cycle}
  \end{center}
\end{figure}

Let $G$ be a combinatorial graph, and $G_2$ the corresponding
two-particle combinatorial  graph.
Let $\epsilon$ and $\phi$ denote disjoint edges of $G$, i.e.~$\epsilon$ and
$\phi$ have no vertices in common.
Let  $c_{\epsilon,\phi} $ denote the cycle on $G_2$ given by
\begin{equation}\label{eq: c_jk,lm}
    c_{\epsilon,\phi} = \left(\{\epsilon_<,\phi_<\}, \{\epsilon_>,\phi_<\}, \{\epsilon_>,\phi_>\},
    \{\epsilon_<,\phi_>\}, \{\epsilon_<,\phi_<\}\right).
\end{equation}
That is, along $c_{\epsilon,\phi}$, the particles move in alternating steps
back and forth along $\epsilon$ and $\phi$.
We say that $c_{\epsilon,\phi}$ is {\it metrically
contractible}.  The reason is that $c_{\epsilon,\phi}$ corresponds to a loop
$\gamma_{\epsilon,\phi}$
on $C_2(\Gamma)$, the two-particle configuration space for a metric graph
$\Gamma$ associated with $G$, which can be continuously contracted to a point
-- see Figure~\ref{fig: contractible cycle}.
(These considerations may
be made more formal.  Let $**$ denote a vertex of the two-particle
configuration space.  One can show that $\pi_1(C_2(\Gamma,**))$ is isomorphic to the
quotient of $\pi^C(G_2,**)$ by the subgroup $K(G_2,**)$ generated by cycles of
the form $p c_{\epsilon,\phi} p^{-1}$, where $\epsilon$ and $\phi$ are disjoint
and $p$ is a path from $**$ to $\{\epsilon_<,\phi_<\}$.)

Let $\Omega_2$ denote a gauge potential on $G_2$. We say that
$\Omega_2$ is a {\it topological
gauge potential} if its flux through every metrically contractible cycle
vanishes modulo $2\pi$, i.e.
\begin{equation}\label{eq: topological gauge potential}
    \Omega_2(c_{\epsilon,\phi}) \equiv 0 \mod 2\pi, \text{ for } \epsilon, \phi
    \text{ disjoint.}
\end{equation}
(More formally,
topological gauge potentials
may be identified with  one-dimensional representations of
$\pi_1(C_2(\Gamma),**)$.)

Informally, gauge potentials that are not topological may be understood to
introduce additional forces into the underlying classical dynamics, as the
following heuristic argument, based on the analogy with metric graphs,
demonstrates: when one particle is on the edge $\epsilon$ and the other on
$\phi$, the free-particle Hamiltonian for the two-particle metric graph is
given by $\half (p^2_\epsilon + p^2_\phi)$. If we introduce a gauge potential,
replacing $p_\epsilon$ by $p_\epsilon - A_\epsilon(x_\epsilon,y_\phi)$ and
$p_\phi$ by $p_\phi - A_\phi(x_\epsilon,y_\phi)$, the equations of motion
become
\begin{equation}\label{eq: equations of motion with vector potential}
    \ddot x_\epsilon = B(x_\epsilon,y_\phi) \dot y_\phi, \quad \ddot y_\phi =
    -B(x_\epsilon,y_\phi) \dot x_\epsilon,
\end{equation}
where the gauge field, $B$, is given by $\partial A_\phi/\partial x_\epsilon -
\partial A_\epsilon/\partial y_\phi$.  If $B \ne 0$ (analogous to
$\Omega_2(c_{\epsilon,\phi}) \ne 0 \mod 2 \pi)$, the classical motion is no
longer free. For combinatorial graphs,  some numerical evidence for the effects
of these
gauge forces is given in
Section~\ref{sec: examples}\ref{subsec: linear}.

\subsection{Characterization of topological gauge potentials}
\label{subsec: characterization}

Given a combinatorial graph $G$, we obtain below an explicit parametrization
of the topological gauge potentials on the two-particle graph, $G_2$.  The
parametrization 
is given in terms of a set of fluxes, or phases. 

Let $T_2$ denote a spanning tree of $G_2$.  As shown in
Section~\ref{sec: one-particle graph}\ref{subsec: gauge
potentials},
given a gauge potential $\Omega_2$ on $G_2$ (not necessarily
topological), we can choose a gauge so that $\Omega_2$ vanishes on the edges of
$T_2$.  Let us label the edges of $G_2$ that are not in $T_2$ by an index
$\phi_2$, and let $c_{\phi_2}$ denote the corresponding fundamental cycles.
The number of such cycles, which we denote by $f_2$, is given by (cf.~\eqref{eq:
f for G})
\begin{equation}\label{eq: g_2}
   f_2 = e_2 - (v_2 - 1) =  e(v-2) - v(v-1)/2 + 1.
\end{equation}
Let $\omega_{\phi_2}$ denote the flux of $\Omega_2$ through $c_{\phi_2}$.
Then $\Omega_2$ is determined by $\omegavec =
(\omega_{1},\ldots,\omega_{f_2})$.

The conditions $\Omega_2(c_{\epsilon,\phi}) = 0 \mod 2\pi $ on topological
gauge potentials (cf.~\eqref{eq: topological gauge potential})
can be expressed as linear relations on the components of $\omegavec$. It will
be convenient to label  pairs of disjoint edges $(\epsilon, \phi)$
by a single index $a$. 
The number of such pairs, which we denote by $g_2$, is just the  difference
between the number of all pairs of edges of $G$ and the number of pairs of
edges that share a vertex, so that
\begin{equation}\label{eq: g - number of disjoint edge pairs}
    g_2 = \half e(e-1) - \half \sum_{j = 1}^v v_j (v_j - 1).
\end{equation}
Therefore, the condition for $\Omega_2$ to be a topological gauge potential can
be written as
\begin{equation}\label{eq: constraint on topological Omega_2}
R\cdot \omegavec = 2\pi \nvec,
\end{equation}
where 
$R$ is a $g_2 \times f_2$ integer matrix and $\nvec$  an arbitrary
$g_2$-dimensional integer vector.  The specific form of $R$
depends on the choice of fundamental cycles.  We note that the
rows of $R$ may be linearly dependent (this turns out to be the
case if, for example, $G$ contains two or more disjoint cycles).

The system \eqref{eq: constraint on topological Omega_2} can be
solved by expressing $R$ in Smith normal form (see, e.g., Dummit
\& Foote 2003).  We write
\begin{equation}\label{eq: Smith}
    R = PDQ,
\end{equation}
where $ P$ and $Q$ are integer matrices with integer inverses of dimensions
$g_2$ and $f_2$ respectively,
and $D$ is a $g_2 \times f_2$ nonnegative diagonal integer matrix. The number
of nonzero diagonal elements of $D$ is given by the rank of $R$, denoted $r$,
and $D$ may be chosen so that its first $r$ diagonal elements are nonzero with
$D_{jj}$ divisible by $D_{j-1,j-1}$ for $1 < j \le r$.  The nonzero diagonal
elements of $D$ are called the {\it divisors} of $R$.  They are uniquely
determined by $R$, and do not depend on the choice of fundamental cycles (i.e.,
they are basis-independent).

Substituting the Smith normal form \eqref{eq: Smith} into \eqref{eq: constraint
on topological Omega_2}, we can write the conditions on $\Omega_2$ as
\begin{equation}\label{eq: contraint on Omega 2 2}
D_{aa} \Phi_a =  2\pi m_a, \quad  1 \le a \le g_2,
\end{equation}
where $\mvec = P^{-1} \nvec$ is integral and $\Phivec = Q \omegavec$.
$\Phi_a$ may be regarded as the flux of $\Omega_2$ through a cycle
$C_a$ in which the fundamental cycle $c_{\phi_2}$ appears  with
multiplicity given by $Q_{a, \phi_2}$.

From \eqref{eq: contraint on Omega 2 2}, the allowed values of
$\Phi_a$ depend on $D_{aa}$, and in particular on whether $D_{aa}$
is one, greater than one, or zero. Let $p$ denote the number of
divisors of $R$ equal to one, $q = r - p$ the number of divisors
greater than one, and  $s = g_2 - r$ the number of vanishing
diagonal elements of $D$.  Then \eqref {eq: contraint on Omega 2
2} can be written as
\begin{align}\label{eq: Phi_p}
     \Phi_j &= 0 \mod 2\pi, &1 \le j \le p,\nonumber\\
     \Phi_{p+ k} &= 2 \pi m_k/d_k \mod 2\pi, \ \ m_k = 0,\ldots,d_k -1, &1 \le k \le
     q,\nonumber\\
     \Phi_{r + l} &= \alpha_l \mod 2\pi,\ \  0 \le \alpha_l < 2 \pi, &1 \le l \le s,
\end{align}
where $d_k = D_{p + k,p +k}$ is shortened notation for the
divisors of $R$ greater than one. Once the $\Phi_a$'s are
specified, $\omegavec$, and hence $\Omega_2$, are determined by
$\omegavec = Q^{-1}\Phivec$.

Thus, topological gauge potentials are parameterized  by $s$ phases
$\alpha_1,\ldots, \alpha_{s}$ taking values between $0$ and $2\pi$, and $q$
phases  $2\pi m_1/d_1, \ldots, 2\pi m_{q}/d_{q}$ taking values constrained to be
rational multiples of $2\pi$.  We shall refer to the $\alpha_l$'s as {\it free
statistics phases} and the $2\pi m_k/d_k$'s as {\it discrete statistics
phases}.  Examples of both are given in Section~\ref{sec: examples}.
(More formally, the set of topological vector potentials modulo trivial gauge
potentials, regarded as a group under matrix addition modulo $2\pi$, is
isomorphic to
 $U(1)^{s} \times \Zz/d_1 \times
\cdots \times \Zz/d_{q}$, and is the character group of
$\pi_1(C_2(\Gamma))$.)

\subsection{Aharonov-Bohm phases and two-body phases}
\label{subsec: AB potentials}

Among the free statistics phases, there may be some that correspond to one of
the particles going around a cycle on $G$ while the other remains fixed.
Physically, such phases could be produced by solenoids threading the cycles of
$G$ with magnetic flux (assuming the particles are charged).  From the point of
view of quantum statistics, we would like to distinguish between contributions
to topological gauge potentials that may be attributed to individual particles
interacting with an external gauge potential, on the one hand, and
contributions involving many-body effects on the other.

Let $\Omega$ be a gauge potential on a combinatorial graph $G$.  We can
construct a corresponding gauge potential on $G_2$, denoted $\Omega^{AB}_2$,
following the prescription \eqref{eq: H^2} for constructing a two-particle
Hamiltonian from a single-particle Hamiltonian, i.e.
\begin{equation}\label{eq: Omega_2 from Omega}
    \Omega^{AB}_{2; \{j,l\}\{k,m\}} = \Omega_{jk} \delta_{lm} +\Omega_{jm} \delta_{kl}
    + \Omega_{lk} \delta_{jm} + \Omega_{lm} \delta_{jk}.
\end{equation}
$\Omega^{AB}_2$ is antisymmetric (since $\Omega$ is antisymmetric), and hence
constitutes a gauge potential on $G_2$.  We shall call gauge potentials of the
form \eqref{eq: Omega_2 from Omega} {\it Aharonov-Bohm potentials}. They are
parameterized (up to a choice of gauge) by  fluxes, or {\it Aharonov-Bohm
phases}, $\phi^{AB}_1,\ldots, \phi^{AB}_f$, through a set of fundamental cycles
on $G$.

Given a cycle $c_2$ on $G_2$, the flux 
$\Omega^{AB}_2(c_2)$ can be evaluated as follows. Let $p(c_2)$ and $q(c_2)$
denote the paths on $G$ described by the individual particles as $c_2$ is
traversed. If each particle returns to where it started, then $p(c_2)$ and
$q(c_2)$ are themselves cycles, and we say that $c_2$ is a {\it direct cycle}.
In this case,  $\Omega^{AB}_2(c_2)$  is the sum of one-particle fluxes, i.e.
\begin{equation}\label{eq: Omega_2 for direct cycle}
    \Omega^{AB}_2(c_2) = \Omega(p(c_2)) + \Omega(q(c_2)).
\end{equation}
Note that \eqref{eq: Omega_2 for direct cycle} implies that $\Omega^{AB}_2$ is
in fact a topological gauge potential;
$p(c_{\epsilon,\phi})$ and $q(c_{\epsilon,\phi})$ are self-retracing cycles on
the edges $\epsilon$ and $\phi$ through which the flux of $\Omega$ obviously
vanishes.
If $c_2$ is not a direct cycle, then the particles exchange positions along
$c_2$, and we say that $c_2$ is an {\it exchange cycle}. In this case, the last
vertex of $p(c_2)$ is the first vertex of $q(c_2)$, and vice versa,
and
\begin{equation}\label{eq: Omega_2 for exchange cycle}
    \Omega^{AB}_2(c_2) = \Omega(p(c_2)q(c_2)).
\end{equation}

We will say that two topological gauge potentials $\Omega_2$ and $\Omega_2'$
are {\it AB-equivalent} if their difference, $\Omega_2 - \Omega_2'$, is an
Aharonov-Bohm potential; equivalently, $\Omega_2'$ can be produced from
$\Omega_2$ by adjusting the Aharonov-Bohm fluxes $\phi^{AB}_j$.   We would like
to find parameters that determine topological gauge potentials up to AB
equivalence.  These parameters, together with the $\phi^{AB}_j$'s, then
provide a complete parametrization of topological gauge potentials (up to a
choice of gauge).

We proceed by fixing a representative $\Omega^*_2$ from each family of
AB-equivalent topological gauge potentials, and then finding parameters that
characterize $\Omega^*_2$.  For the special case of circular graphs this is
easily done.  Circular graphs have their vertices arranged on a ring, and the
only edges are between adjacent vertices.  We may take $\Omega^*_2 = 0$,
because every topological gauge potential $\Omega_2$ may be generated by an Aharonov-Bohm flux
through the ring.   Circular graphs are discussed further in
Section~\ref{sec: examples}\ref{subsec: circular}.

Assuming that $G$ is not a circular graph, we can choose a set of fundamental
cycles $c_j$ on $G$ such that none of the $c_j$'s contains every
vertex of $G$.  Let $k_j$ denote a vertex of $G$ not contained in $c_j$, and
let $c_{2; j}$ denote the cycle on $G_2$ on which one particle traverses
$c_{j}$ while the other remains fixed at $k_j$.   We fix $\Omega_2^*$  (up to
a choice of gauge) by  requiring that
\begin{equation}\label{eq: fix AB gauge}
    \Omega_2^*(c_{2; j}) = 0 \mod 2\pi
\end{equation}
(a straightforward argument shows that every topological gauge potential is $AB$-equivalent to a
unique topological gauge potential satisfying \eqref{eq: fix AB gauge}).

The $f$ linear conditions \eqref{eq: fix AB gauge} on $\Omega^*_2$
can be combined with the $g_2$ linear conditions \eqref{eq:
topological gauge potential} which characterize general
topological vector potentials, and the two sets of conditions
written collectively as (cf.~\eqref{eq: constraint on topological
Omega_2})
\begin{equation}\label{eq: augmented relations}
    R^* \cdot \omegavec^* = 2\pi \nvec^*,
\end{equation}
where $R^*$ is an integer matrix of dimension $(f+g_2) \times f_2$, and
$\nvec^*$ is an integer vector of dimension $(f+g_2)$.  The solution of \eqref{eq:
augmented relations} proceeds as in
Section~\ref{sec: 2 particles}\ref{subsec: characterization}.
Solutions are parameterized by $s-f$ phases $\beta_1,\ldots,
\beta_{s-f}$, which we call {\it two-body phases}, and  $q$
discrete statistics phases $2\pi m_1/d_1, \ldots, 2\pi
m_{q}/d_{q}$.  It follows that a general topological gauge
potential may be parameterized by $f$ Aharonov-Bohm phases
$\phi^{AB}_1, \ldots, \phi^{AB}_f$, which determine its fluxes
through the $c_{2;j}$'s, in addition to the  two-body phases
$\beta_l$ and discrete statistics phases $2\pi m_k/d_k$.  The
lasso and bowtie graphs
(Section~\ref{sec: examples}\ref{subsec: lasso and
figure-of-eight})
provide simple illustrations of the distinction between
Aharonov-Bohm and two-body phases.

\subsection{Bose and Fermi statistics}\label{subsec: Bose and Fermi}

Our treatment of quantum statistics on combinatorial graphs  -- referred to in what follows as the {\it identified
scheme} --
follows the approach of Leinaas and Myrheim (1977),
treating the particles as classically indistinguishable.  If instead one
followed the conventional approach -- referred to in what follows as the {\it
distinguished scheme}, in which the particles are labeled from the start, one would find
Bose and Fermi statistics.  Below we establish the relationship
between the two schemes.  Note that, in the identified scheme, it does not make
sense to speak of the exchange symmetry of a  quantum state, as exchange is
not defined -- the wavefunction assumes a single value for each configuration of the
indistinguishable particles.  It turns out that Bose and Fermi statistics may
be regarded as
particular cases of the more general quantum statistics we have obtained
here. 
In particular,
Bose statistics corresponds to a trivial (e.g., vanishing) topological gauge
potential, while Fermi statistics corresponds to a topological gauge potential
that assigns a phase of $\pi$ to  exchange cycles and zero phase to direct cycles.

Following the distinguished scheme one proceeds as follows.
Let $G$ denote a combinatorial graph with adjacency matrix $A$.
We introduce the  distinguished two-particle configuration space, denoted
$\Cbar_2(G)$, or $\Gbar_2$ for short, that
consists of ordered pairs $(j,l)$, $j \ne l$, of distinct
vertices of $G$ (note that particles are still not allowed to occupy the
same vertex).  We regard $\Gbar_2$ as
a graph with $v(v-1)$ vertices and adjacency matrix
\begin{equation}\label{eq: A^2D}
    \Abar_{2; jl,km} = A_{jk} \delta_{lm} + A_{lm} \delta_{jk}.
\end{equation}
Following the general quantization prescription given in
Section~\ref{sec: one-particle graph}\ref{subsec: 1-particle
quantization},
we introduce the distinguished Hilbert space
$\Hcalbar_2 = \Cc^{v(v-1)}$, with basis vectors $\ket{\overline{jl}}$, $j \ne
l$, which
 describe states where particle 1 is at vertex $j$ and
particle 2 at vertex $l$. A quantum Hamiltonian on $\Hcalbar_2$, denoted $\Hbar_2$, is a
$v(v-1)$-dimensional matrix with matrix elements
\begin{equation}\label{eq: H^D}
    \Hbar_{2; jl,km} = \bra{\overline{jl}} \Hbar_2 \ket{\overline{km}}.
\end{equation}
satisfying \eqref{eq: constraint on H}  (so that short-time
transitions are single-particle transitions).

Exchange is defined on $\Hcalbar_2$ by $\ket{\overline{jl}}
\rightarrow \ket{\overline{lj}}$.
We decompose $\Hcalbar_2$  into two
$v(v-1)/2$-dimensional subspaces, $\Hcalbar_{2}^{\sigma}$,  consisting of
states
that are even ($\sigma = 1$) or odd ($\sigma = -1$) under exchange.
Indistinguishability is incorporated by requiring the Hamiltonian (and, indeed,
any hermitian matrix representing an observable) to be invariant under
exchange, i.e.~$\Hbar_{2; jl, km} = \Hbar_{2; lj, mk}$.
Then $\Hbar_2$ preserves exchange symmetry, i.e.~it leaves the subspaces
$\Hcalbar^\sigma$ invariant.
%

For example, given a one-particle Hamiltonian $H$, we may construct the two-particle Hamiltonian 
\begin{equation}\label{eq: H^D as sum of one-particle Hamiltonians}
    \Hbar_{2; jl, km} = H_{jk} \delta_{lm} + H_{lm} \delta_{jk}.
\end{equation}
On the antisymmetric subspace $\Hcal_2^-$,
the energy levels of $\Hbar_2$ are just sums of
distinct energy levels of $H$, and the two-particle eigenstates are
antisymmetric products of one-particle eigenstates -- this corresponds to
noninteracting fermions.
The situation is different
on the symmetric subspace
$\Hcalbar_2^+$.  In general, the spectrum of $H_2$ on $\Hcalbar_2^+$
is not simply related to the one-particle spectrum (although in
some special cases it is).  This is because the diagonal states
$\ket{\overline{jj}}$ are excluded; bosons are
prevented from occupying the same vertex, which, unlike fermions, they would
otherwise do.

The equivalence between the identified and distinguished schemes
may be established by means of a unitary map, $U^\sigma$, from the
identified Hilbert space, $\Hcal_2$, to the symmetric or
antisymmetric subspace, $\Hcalbar_2^\sigma$, of the distinguished
Hilbert space.
$U^\sigma$ is given by
\begin{equation}\label{eq: U}
    U^\sigma \ket{jl} = \frac{1}{\sqrt 2} \left(  \ket{\overline{jl}} + \sigma
    \ket{\overline{lj}}\right), \ \ j < l.
\end{equation}
Given an exchange-invariant Hamiltonian $\Hbar_2$ on
$\Hcalbar^\sigma$, we can construct a
unitarily equivalent Hamiltonian $H_2^\sigma$ on
$\Hcal_2$ by taking
$H_2^{\sigma} = U^\sigma \Hbar_2 {U^\sigma}^\dag$. The matrix elements of $H_2^{\sigma}$
are given explicitly by
\begin{equation}\label{eq: translation of H^2}
    H^{\sigma}_{2; jl,km} =  \Hbar_{2; jl,km} + \sigma \Hbar_{2; lj,km}, \ \ j
    < l, k < m.
\end{equation}
Thus, in the identified scheme, the statistics sign $\sigma$ appears in
the Hamiltonian $H_2^\sigma$, while the Hilbert space, $\Hcal_2$, is
independent of $\sigma$.  In contrast, in the distinguished scheme, $\sigma$ determines
the Hilbert space $\Hcalbar_2^\sigma$, while the
Hamiltonian $\Hbar_2$ is independent of $\sigma$.
In particular, if $\Hbar_2$ is given by the sum of one-particle Hamiltonians as
in \eqref{eq: H^D as sum of one-particle Hamiltonians}, then \eqref{eq: translation of
H^2} yields
\begin{equation}\label{eq: H^2 second}
    \bra{jl} H^\sigma_2 \ket{km} = \delta_{jk} H_{lm}
    + \sigma \delta_{jm} H_{lk} + \sigma \delta_{lk} H_{jm} +
    \delta_{lm} H_{jk}, \quad j < l, k < m.
\end{equation}
This is in accord with \eqref{eq: H^2}, and establishes
that $\sigma$ corresponds to Bose or Fermi statistics.

From \eqref{eq: translation of H^2} and the constraints \eqref{eq:
constraint on H},  we can express the relation between the Bose
and Fermi Hamiltonians $H^{-}_2$ and $H^{+}_2$ as
\begin{equation}\label{eq: H_+ and H_- second}
H^{-}_{2; jl,km} = e^{i\Omega^{F}_{2; jl,km}}H^{+}_{2; jl,km},
\end{equation}
where $\Omega^{F}_2$ is given by
\begin{equation}\label{eq: Omega Bose Fermi}
    \Omega^{F}_{2; jl,km} =
\begin{cases}
\pi,& j = m, l \sim k ,\\
-\pi,&j \sim m, l = k,\\
0,& \text{otherwise}.
\end{cases}
\end{equation}
Thus,
$H^+_2$ and $H^-_2$ are related by a gauge potential, as
$\Omega^{F}_2$ is real antisymmetric and its ${(jl,km)}$th element
vanishes unless $\{j,l\} \sim \{k,m\}$.
In fact, 
$\Omega^{F}_2$ is a topological gauge potential.  This can be
verified by checking the conditions \eqref{eq: topological gauge
potential} explicitly. Alternatively, we may argue as follows:
given a path $p_2$ on $G_2$ of length $r$, let $p(p_2) =
(x_1,\ldots,x_r)$ and $ q_2(p) = (y_,\ldots,y_r)$ denote the
single-particle paths along $p_2$.  Nonzero contributions of $\pm
\pi$ to $\Omega^{F}_2(p_2)$  come from those edges of $p_2$ where
the ordering of the single-particle positions switches over,
either from $x_j < y_j$ to $x_{j+1} > y_{j+1}$ or from $x_j > y_j$
to $x_{j+1} < y_{j+1}$.  However, on a direct cycle $c_2$, and, in
particular, on  $c_{\epsilon,\phi}$, the particles separately
return to where they started, so the number of switchovers, and
hence the number of $\pm \pi$ contributions, must be even --
therefore $\Omega^{F}_2(c_2)$ is a multiple of $2\pi$. Indeed, by
this reasoning we see that
\begin{equation}\label{eq: Omega Bose Fermi on cycles}
    e^{i\Omega^{F}_2(c_2)} = \begin{cases}
    1, & \text{if $c_2$ is a direct cycle},\\
    -1,& \text{if $c_2$ is an exchange cycle}.    \end{cases}
\end{equation}

\section{Examples}\label{sec: examples}
We investigate statistics phases for a number of graphs,
starting with the simplest examples.

\subsection{Linear graphs}\label{subsec: linear}
The linear graph $L_N$ consists of $N$ vertices on a line
with adjacent vertices connected by edges,
so that $A_{j,k} = \delta_{|j-k|,1}$.
From the point of view of quantum statistics, linear graphs are trivial; it turns out
that there are no nontrivial topological gauge potentials on
two-particle linear graphs.
However, linear graphs provide simple examples where the free-particle energy levels and eigenstates
can be calculated explicitly, and they serve to illustrate
some points of the preceding discussion, including the equivalence of Bose and Fermi
statistics in cases where topological phases are absent, as well as the effect of non-topological gauge
potentials.

Let $H$ be the one-particle kinetic energy on $L_N$ (cf.
Section~\ref{sec: one-particle graph}\ref{subsec: 1-particle
quantization}.
It is straightforward to show that
the energy levels of $H$ are given by $E_a = 4\sin^2(\pi a/2N)$,
$0 \le a < N$, while the eigenstates are given, up to
normalization, by $\braket{j}{\psi_a} = \cos(\pi a j/N)$. We take
the two-particle Hamiltonian, $H_2^\sigma$, to be the sum of
one-particle Hamiltonians as given by \eqref{eq: H^2}.  As
discussed in
Section~\ref{sec: 2 particles}\ref{subsec: Bose and Fermi},
$\sigma =  1$ corresponds to Bose statistics and $\sigma=-1$ to
Fermi statistics. It is straightforward to show that the energy
levels of $H_2^\sigma$ are independent of $\sigma$, and are given
by sums $E_a + E_b$ with $a$ and $b$ distinct.  The corresponding
eigenstates are, for $\sigma = -1$, antisymmetric products of
distinct eigenstates $\ket{\psi_a}$ and $\ket{\psi_b}$, and, for
$\sigma = +1$, symmetric products of these eigenstates with
diagonal components removed. The interaction between the
(hard-core) bosons is reflected in the fact that two-particle
states with $a = b$ are absent from the spectrum.

The fact that the Bose and Fermi spectra coincide for linear graphs can be understood from
our
topological treatment of quantum statistics.
The only nontrivial cycles on the two-particle configuration space
correspond to each particle moving back and
forth in alternation along separated segments of the linear graph. Such cycles
are metrically contractible, so that there
are no nontrivial topological gauge potentials.  It follows that Fermi gauge
potential $\Omega^{F}_2$ (cf.~\eqref{eq: Omega Bose Fermi})
is just a gauge transformation.

There is an alternative, single-particle interpretation of $H^\sigma_2$,
namely  as a
discrete approximation to the quantum Hamiltonian for a free particle in
a two-dimensional right triangular domain
with sides $1$, $1$ and $\sqrt 2$ (i.e., half the unit square
below the diagonal), and with $\hbar = 1/N$.  Neumann boundary conditions apply on the right sides
of the triangle, while $\sigma$ determines the boundary condition on the diagonal
(Neumann for $\sigma = 1$, Dirichlet for $\sigma = -1$).

\begin{figure}[htb]
\centering
\subfigure[]
{    \label{fig:states:a}
    \includegraphics[width=6cm]{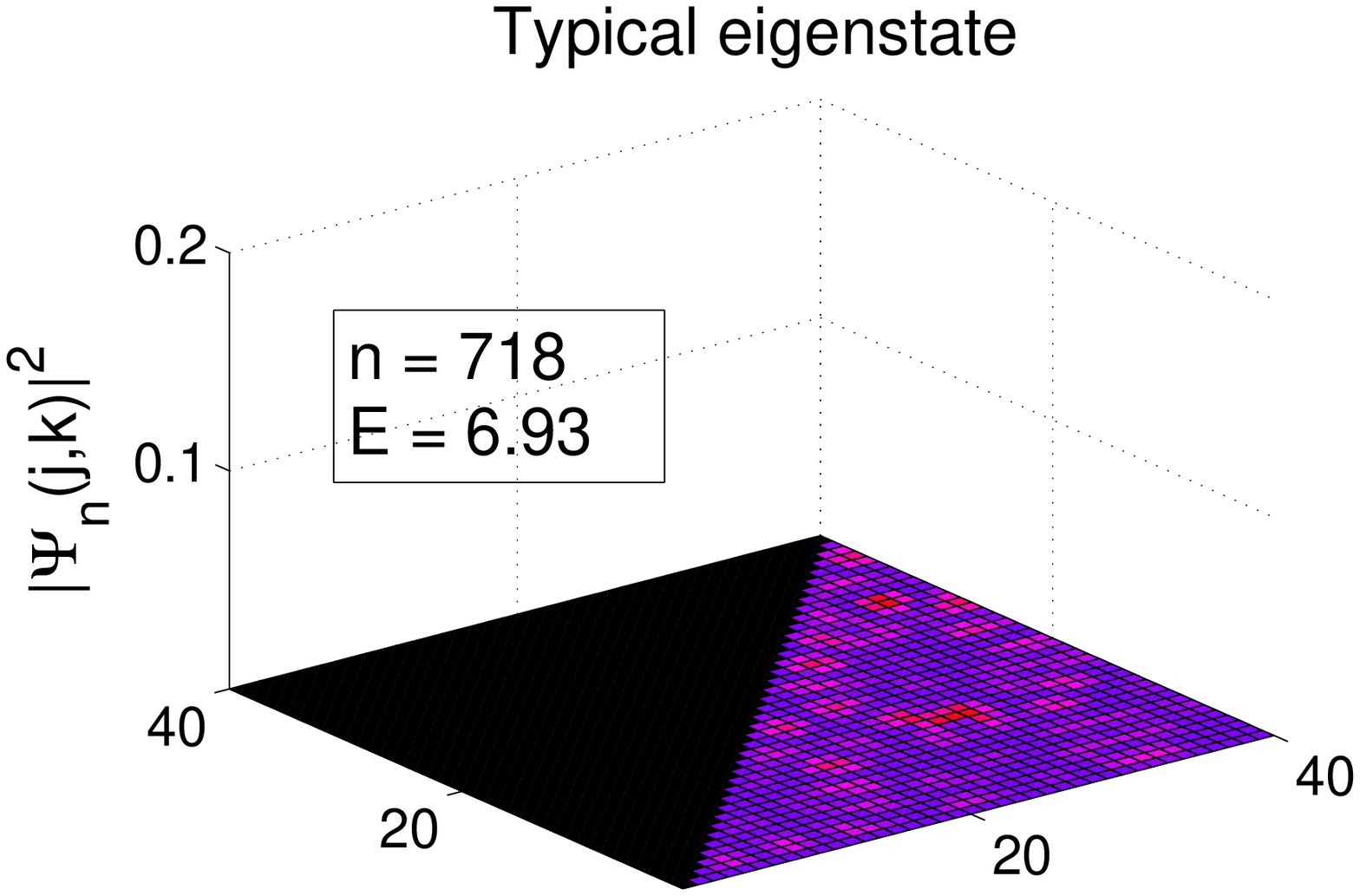}}
\subfigure[]
{
    \label{fig:states:b}
    \includegraphics[width=6cm]{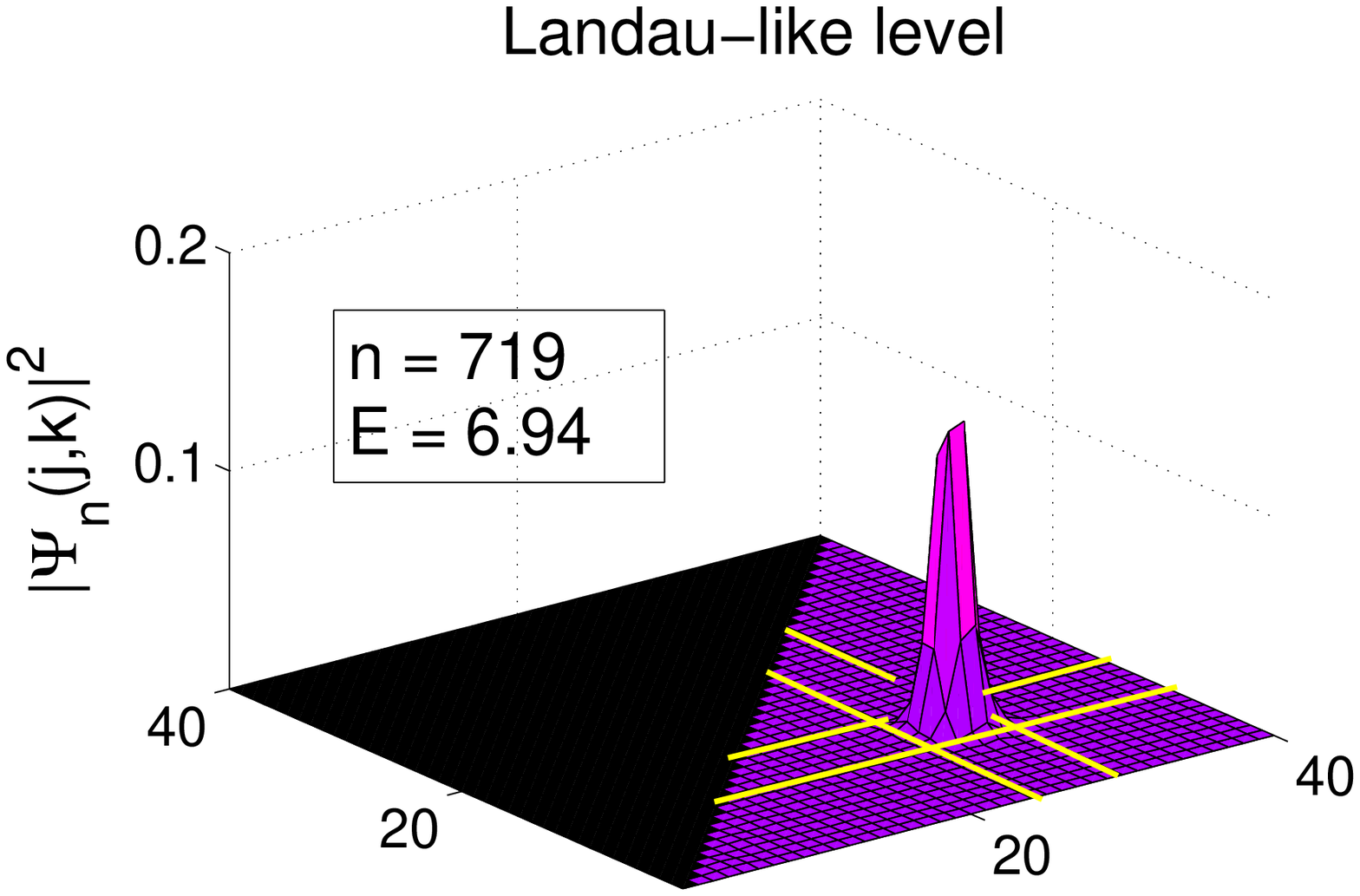}}
\caption{Consecutive eigenstates for two particles on a linear
graph with $N=40$ vertices with gauge potential (a) Typical
eigenstate (b) Eigenstate localized in the flux square, an
analogue of a Landau-level wavefunction} \label{fig:states}
\end{figure}

Fixing the Hamiltonian to be $H_2^-$ for definiteness, it is
instructive to observe the consequences of introducing a gauge
potential, denoted $\Omegahat_2$, that is {\it not} topological.
Let us take $\Omegahat_2$ to have flux $2\pi p/t$ through every
cycle $c_{\epsilon,\phi}$ where $\epsilon$ is an edge between
vertices in the range between $r$ and $r +t$, and  $\phi$  an edge
between vertices in the range between $s$ and $s+t$.  For cycles
$c_{\epsilon,\phi}$ outside this range, we take
$\Omegahat_2(c_{\epsilon,\phi}) = 0$. In terms of the
single-particle interpretation, $\Omegahat_2$ corresponds to a
uniform magnetic field of strength $2\pi p /(t/N)$ (in units where
$q/c$,  $q$ being the charge of the particle, is equal to one)
through the square of area $t^2/N^2$ with diagonal corners $(r/N,
s/N)$ and $((r+t)/N,(s+t)/N)$. Numerical calculations show that
the shifts in energy levels produced by $\Omegahat_2$
scale with $1/N$, in accord with the fact that a magnetic field
does not change the mean density of states. However, with
$\Omegahat_2$ present, while typical eigenstates are delocalized
(cf.~Figure~\ref{fig:states:a}), one finds some eigenstates that
are strongly localized in the flux square $10 \le j \le 15$ and
$25 \le l \le 30$ (cf.~Figure~\ref{fig:states:b}).  In the
single-particle interpretation these are Landau-like levels, and
are indicative of Lorentz forces in the classical dynamics.

\subsection{Circular graphs}\label{subsec: circular}

\begin{figure}[htb]
\begin{center}
\setlength{\unitlength}{1cm}
\begin{picture}(8,2.5)
\put(0.5,0.3){\includegraphics[width=2.5cm]{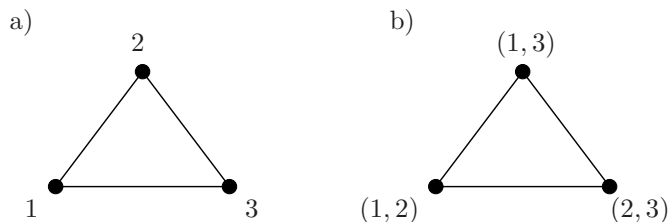}}
\put(5.5,0.3){\includegraphics[width=2.5cm]{triangle.eps}} \put(0,2.5){a)}
\put(5,2.5){b)} \put(0.2,0){$1$} \put(3.1,0){$3$} \put(1.6,2.2){$2$}
\put(4.6,0){$(1,2)$} \put(7.9,0){$(2,3)$} \put(6.4,2.2){$(1,3)$}
\end{picture}
\caption{(a) $C_3$ (b) Two-particle graph for
$C_3$}\label{fig:triangle}
\end{center}
\end{figure}

The circular graph, $C_N$, is obtained by connecting the first and last vertices of the linear graph
$L_N$.   For a circular graph, we naturally expect, and indeed find, anyon
statistics, confirming that our model provides a reasonable description of quantum
mechanics on a loop.  For simplicity we consider a loop with three vertices, shown
in Figure~\ref{fig:triangle}~(a) (the conclusions are similar for $N>3$).
Writing down the three allowed two-particle configurations  in
Figure \ref{fig:triangle}~(b),  it is apparent that the two-particle graph $G_2$
is also a single loop.  A single traversal of this loop is an exchange cycle, and the
associated flux $\phi$ corresponds to anyon
statistics.
The flux $\phi$ can be generated by an Aharonov-Bohm potential,
corresponding to a solenoid threading the one-particle graph $C_3$.

\subsection{Star graphs}\label{subsec: star}

\begin{figure}[htb]
\begin{center}
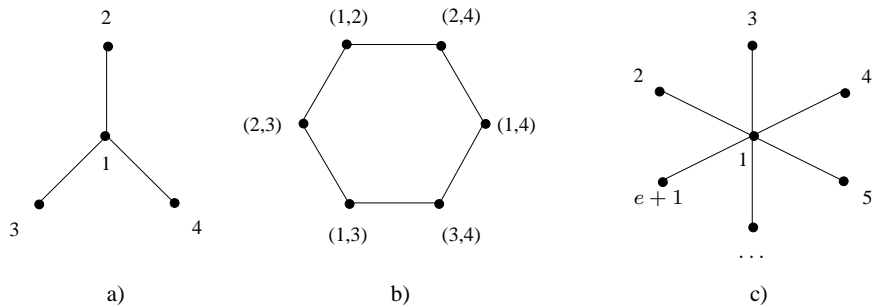
\caption{(a) Y-graph (b) Two-particle Y-graph
(c) Star graph with $e$ edges}\label{fig:stargraph}
\end{center}
\end{figure}

The star graph, $S_e$, shown in Figure~\ref{fig:stargraph}~(c),
consists of $e$ vertices each connected to a central vertex, and
so has $e$ edges and $e+1$ vertices.  We consider first the $e =
3$ star graph, or $Y$-graph, for which the two-particle graph is
easily displayed (Figures~\ref{fig:stargraph} (a) and (b)). The
two-particle graph consists of a single cycle which exchanges the
particles through the arms of the `Y'. A flux through this cycle
produces anyon statistics. For $e > 3$, the two-particle star
graph consists of $v_2 = (e+1)e/2$ vertices and $e_2 = e(e-1)$
edges.  The number of independent cycles, $f_2$, is given by
$(e-1)(e-2)/2$.  There are no Aharonov-Bohm phases (there are no
nontrivial cycles on $S_e$) and no constraints on topological
gauge potentials (there are no disjoint edges on $S_e$).
Therefore, topological gauge potentials on $S_e$ are parameterized
by $g_2 = (e-1)(e-2)/2$ two-body statistics phases.  The number of
statistics phases can also be obtained from the following simple
argument:  each phase corresponds to the choice of a pair of edges
along which to exchange the particles, given that the particles
start on the vertices of some given edge. $g_2$ is therefore the
number of pairs of $e-1$ objects.

\subsection{Lasso and bowtie}\label{subsec: lasso and figure-of-eight}

\begin{figure}[htb]
\begin{center}
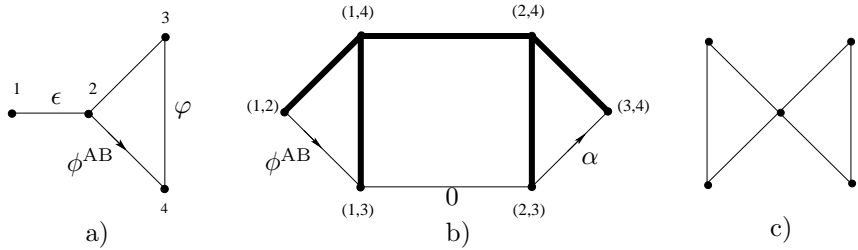
\caption{(a) Lasso (b) Two-particle lasso
(c) Bowtie}\label{fig:lasso+bowtie}
\end{center}
\end{figure}

The lasso graph consists of a three-vertex loop with a single external
lead -- see Figure~\ref{fig:lasso+bowtie}~(a).  It provides a simple example of quantum statistics
which
combines aspects of circular graphs and star graphs discussed above. The two-particle
lasso, $G_2$,
is shown Figure~\ref{fig:lasso+bowtie}~(b) along with a spanning tree, $T_2$ (in bold).
We fix the gauge
so that the edges of $T_2$ are assigned zero phase.
The three edges of $G_2$ not in $T_2$  determine the fundamental
cycles.  The central square corresponds to the metrically contractible
cycle $c_{\epsilon,\varphi}$
 on which the particles move in alternation along the edges $\epsilon$ and $\varphi$
of the lasso. As the flux through this cycle must vanish, the edge $((1,3),(2,3))$ is
assigned zero phase as well.
The left triangle of $G_2$ corresponds to the cycle in which the one of the particles
goes around the loop of the lasso while the other remains on the external lead,
and may be assigned an Aharonov-Bohm phase $\phi^{AB}$.
The right triangle of $G_2$ corresponds to an exchange cycle in which the
particles move around the loop of the lasso.  The associated phase, denoted
$\alpha$, is a two-body phase.  The length-six cycle along the perimeter of $G_2$ coincides
with the exchange cycle on the $Y$ graph, and has phase $\alpha + \phi_{AB}$.
These results coincide with those of Balachandran \& Ercolessi (1992), who considered the two-particle
metric lasso graph.

Another example considered by Balachandran \& Ercolessi (1992) is the bowtie, or figure-of-eight, which consists of two
three-vertex loops sharing a common vertex  -- see Figure~\ref{fig:lasso+bowtie}~(c).  Calculations show
that the two-particle bowtie has two Aharonov-Bohm phases (corresponding to the
two loops) and two two-body phases.


\subsection{Nonplanar graphs: $K_5$, $K_{3,3}$ and the $K_5$ molecule}\label{subsec: K_5 and K_3,3}
$K_5$ is the fully connected graph with five vertices, and
$K_{3,3}$ is the fully connected bi-partite graph
with two sets of three vertices -- see Figure \ref{fig:
 K5K33K5K5}~(a) and (b).
A theorem of Kuratowski (1930)
states that every non-planar graph (i.e., a graph that cannot be
drawn in the plane without crossings) contains $K_5$ or $K_{3,3}$ as a
subgraph (possibly after contracting some edges to points).  From the point of view of
quantum statistics, $K_5$ and $K_{3,3}$ are interesting because they
are the smallest graphs that exhibit discrete statistics phases.
Calculations following the procedure of
Section~\ref{sec: 2 particles}\ref{subsec: characterization}
(details are omitted) show that $K_5$ has
six Aharonov-Bohm phases and $K_{3,3}$ has four Aharonov-Bohm phases
(corresponding to their respective number of fundamental cycles).  In addition,
both have a single discrete statistics phase that can be either $0$ or $\pi$ (mod $2\pi$).
Cycles whose fluxes are given by this discrete phase
are necessarily exchange cycles.  An example for $K_5$ consists of
the cycle in which one particle goes around a three-vertex loop
(e.g., $(1,2,3,1)$) while the other remains fixed, followed by an
exchange around the same three-vertex loop in the opposite
direction (e.g., $((1,2),(2,3),(1,3),(1,2))$). An example for
$K_{3,3}$ consists of the cycle in which one particle goes around
a four-vertex bowtie loop (e.g., $(1,4,2,5,1)$) while the other
remains fixed, followed by an exchange around the same four-vertex
loop in the opposite direction (e.g.,
$((1,4),(4,5),(1,5),(1,2),(1,4))$).

\begin{figure}
\begin{center}
  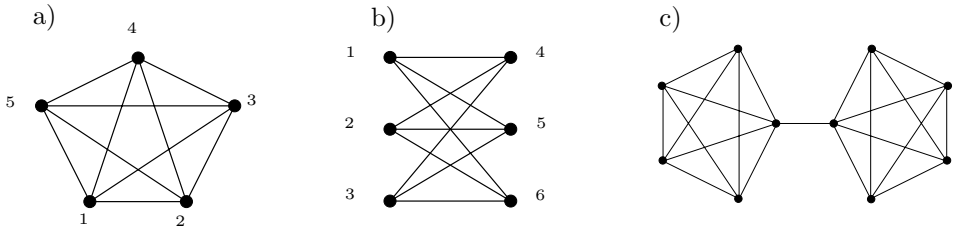
  \caption{The basic non-planar graphs (a) $K_5$ and (b)
$K_{3,3}$, and (c) the $K_5$ molecule.}\label{fig: K5K33K5K5}
  \end{center}
\end{figure}

The $K_5$ molecule is the graph consisting of two $K_5$'s joined by a
single edge, as in Figure~\ref{fig: K5K33K5K5}~(c).  Calculations (again, details are omitted)
show that two discrete
statistics phases appear, both either $0$ or $\pi$, along with 12 Aharonov-Bohm phases
and 6 two-body phases.

\section{Discussion}\label{sec: discussion}

The abelian statistics of two indistinguishable quantum particles
on a combinatorial graph are characterized by a set of continuous
and discrete-valued phases.  The continuous phases may be
separated into some that are produced by external Aharonov-Bohm fluxes and
the rest describing two-body statistical interactions.  The
appearance of discrete phases may be related to whether the graph
is planar or not, a connection that merits further investigation. While we
have concentrated on the case of two particles, the abelian
statistics for more than two particles may be characterized and
calculated using the results presented here.

Nonabelian statistics requires new considerations.  A full
description is needed of the fundamental group of the $n$-particle
configuration space, or, equivalently, the braid groups of the
graph (not simply their abelianized versions). Recently, Farley \&
Sabalka (2005) have developed methods based on discrete Morse
theory for obtaining efficient presentations of graph braid
groups.  We will discuss applications of these techniques to
nonabelian graph statistics in a forthcoming publication (see
(Kitaev 2008) for an exact solution of a spin-lattice model where
nonabelian statistics emerge).

From the point of view of physics, one of the principal
attractions of quantum graphs is that they provide mathematically
tractable models of complex physical systems.  As applications
 have so far concentrated on independent-particle models, the scope
for manifestations of quantum statistics on graphs is great. We
suggest a few possibilities here. Graph statistics may play a role
in many-electron network models of molecules, in analogy with the
emergence of the Berry phase -- the molecular Aharonov-Bohm effect
(Mead and Truhlar 1979) -- in molecular spectra. Topological
signatures in single-particle transport on networks (Avron 1995),
which provide models and variants of the quantum Hall
effect, may have many-particle generalizations in which statistics
plays a role.  Many-particle graphs may provide
new models for anyon superconductivity (Wilczek 1990).
An intriguing potential application of nonabelian
graph statistics is to topological quantum computing (Nayak {\it et al}.~2008).
 There one looks for systems with a degenerate ground state
spanned by distinct quasiparticle configurations, in which the
only (easily) realizable evolutions are, up to a phase, a
discrete set of unitary transformations generated by the
(adiabatic) exchange of  quasiparticles.  By introducing spin
(Harrison 2008), quantum graphs might also provide models in which to
investigate the role of quantum statistics in the quantum spin
Hall effect and topological insulators (Hasan \& Kane 2010, Qi \&
Zheng 2010).

Applications will depend on nontrivial graph statistics emerging
in a particular model.  A standard approach would be to look for
novel many-particle ground states on graphs and to study their
excitations. This is work for the future. However, even without a
specific mechanism, we believe it is worthwhile to pursue a
general investigation of
 quantum statistics on networks and its
consequences. Quantum physics has provided many examples where
topology underlies new and unexpected phenomena. Nature seems to
exploit the opportunities available to it, and discoveries may
follow from knowing where to look and what to look for.
\\

\noindent {\it Acknowledgements.}
We thank Dan Farley for helpful discussions.  JMH is supported by National
Science Foundation grant DMS-0604859.

\label{lastpage}

\end{document}